\documentclass[10pt,paperletter]{article}
\usepackage{graphicx}
\usepackage{amssymb}
\usepackage{amsmath}
\usepackage{latexsym}

\title{Remarks about decay widths of two-body nonleptonic and
semileptonic $B$ decays}

\author{J. H. Mu\~noz\thanks{jhmunoz@ut.edu.co}  $\;$ and N. Quintero\thanks{nquinte@gmail.com}\\Departamento de F\'isica, Universidad del Tolima\\
Apartado A\'ereo 546, Ibagu\'e, Colombia}

\begin{document}
\maketitle

\begin{abstract}
We present a brief discussion about  expressions of decay widths of exclusive nonleptonic and semileptonic $B$ decays at tree level including
$l=0$ and $l=1$ mesons in  final state.  Our analysis is carried out assuming  factorization hypothesis and using parametrizations of hadronic matrix elements given in WSB and  ISGW  quark models.
Special interest is focused on  dynamics of these processes and several important ratios between decay widths to determine form factors and
decay constants   are given.
\end{abstract}

\noindent
{\bf PACS numbers(s):} 13.20.-v, 13.25.-k, 12.39.-x

\vspace{0.3cm}

\newpage

\section{Introduction}

Exclusive semileptonic and nonleptonic $B$ decays offer a  good scenario  for studying, at theoretical and experimental levels,  CP violation, precise determination of Cabibbo-Kobayashi-Maskawa (CKM) matrix elements, new physics beyond the Standard Model, QCD and electroweak penguin effects, production of orbitally excited mesons, etc, which are topics   of great interest of research in particle physics of this decade \cite{updated}.  \\

The purpose of this paper is to perform an overview about expressions of decay widths of exclusive nonleptonic and semileptonic $B$ decays at tree level including $l=0$ and $l=1$ mesons in  final state. For $l=0$, we have considered pseudoscalar ($P$) and vector ($V$) mesons, and for $p$-wave mesons we have included scalar ($S$), axial ($A$) and tensor ($T$) mesons (see Table I). We
present an interesting summary and a panoramic analysis about  expressions of decay
widths of nonleptonic $B \rightarrow
P_{1}P_{2},VP,V_{1}V_{2},AP,AV, A_{1}A_{2}, PS,$ $SS, SV, SA, TP,TV, TS, TA$ decays\footnote{For
definiteness we use a $B$-meson in our notation, but the results
are quite general. They apply equally well to $D$-mesons, or even
to pseudoscalar mesons.}, and  differential decay rates of  semileptonic $B \rightarrow (P,V,S$, $A,T)l\nu$ decays. We have assumed  factorization hypothesis and used  parametrizations of  hadronic matrix element $\langle M|J_{\mu}|B \rangle$ given in  relativistic WSB \cite{wsb} and  nonrelativistic ISGW \cite{isgw} quark models. \\

Our aim is to give a general point of view about $\Gamma(B
\rightarrow M_{1}M_{2})$ and $d\Gamma(B \rightarrow Ml\nu)/dt$,
illustrating  the dynamics of these decays
and pointing out that some relations among these expressions and
form factors that participate in the parametrization of
hadronic matrix element $\langle M| J_{\mu}|B\rangle$, which is common in both
processes, could be experimental tests. For the sake of simplicity
we have only considered tree level color-allowed external W-emission
nonleptonic $B$ decays, i.e. the so called \emph{class-I} decays in the
literature \cite{neubert}.\\

The paper is organized  as follows:  Section 2 contains
expressions for $\Gamma(B \rightarrow M_{1}M_{2})$ and $d\Gamma(B
\rightarrow Ml\nu)/dt$ and a discussion about them. In section 3, we analyze vector and axial contributions of weak interaction to $B \rightarrow (P,V,S,A,T)l\nu$ decays assuming a meson dominance model.  In section 4, we quote some important relations among decay widths, which allow us to determine some form factors and decay constants.  Finally, in section 5 we present some concluding remarks.

\begin{table}[ht]
{\small Table I. Spectroscopic notation for $l=0$ and $l=1$ mesons. $l$ is the orbital angular momentum, $s$ is the spin, and $J$ is the total angular momentum. $P$ and $C$ are  parity and  charge conjugate operators, respectively. }
\par
\begin{center}
\renewcommand{\arraystretch}{1.5}
\begin{tabular}{|c|c|c|c|c|c|}
  \hline
  $l$ & $s$ & $J$ & $^{2s+1}L_{J}$ & $J^{PC}$ & Meson \\
  \hline\hline
  $l=0$ & $s=0$ & $J=0$ & ${}^{1}S_{0}$ & $0^{-+}$ & Pseudoscalar $(P)$ \\
  \cline{2-6}
    & $s=1$ & $J=1$ & ${}^{3}S_{1}$ & $1^{--}$ & Vector $(V)$ \\
    \hline
    & $s=0$ & $J=1$ & ${}^{1}P_{1}$ & $1^{+-}$ & Axial-vector $\left(A({}^{1}P_{1})\right)$ \\
    \cline{2-6}
  $l=1$ &   & $J=0$ & ${}^{3}P_{0}$ & $0^{++}$ & Scalar $(S)$ \\
  \cline{3-6}
    & $s=1$ & $J=1$ & ${}^{3}P_{1}$ & $1^{++}$ & Axial-vector $\left(A({}^{3}P_{1})\right)$ \\
    \cline{3-6}
    &   & $J=2$ & ${}^{3}P_{2}$ & $2^{++}$ & Tensor $(T)$ \\
  \hline
\end{tabular}
\end{center}
\end{table}

\section{$\Gamma(B \rightarrow M_{1}M_{2})$ and $d\Gamma(B
\rightarrow Ml\nu)/dt$}

In this section we summarize  expressions at tree level of
differential decay rates of $B \rightarrow (P,V,S,A,T)l\nu_{l}$
(see Table II) and  decay widths of $B(b\bar{q}') \rightarrow
M_{1}(q\bar{q}')M_{2}(q_{i}\bar{q}_{j})$, where
$M_{1,2}$ can be a pseudoscalar $(P)$, a vector $(V)$, a scalar $(S)$, an axial-vector $(A)$ or a tensor
$(T)$ meson (see Table III)\footnote{The Ref. \cite{korner} also gives a summary about
differential decay rates of $\Gamma(B \rightarrow P(V)l\nu_{l})$
and  decay rates of $B \rightarrow
P_{1}P_{2},PV,VP,V_{1}V_{2}$.}. For  nonleptonic two-body $B$ decays we have used
the notation $B \rightarrow M_{1},M_{2}$ \cite{chen} to mean that
$M_{2}$ meson is factorized out under   factorization approximation,  i.e., $M_2$ arises from  vacuum.  For  $B \rightarrow P$ and $B \rightarrow V$ transitions
we have used the parametrizations given in  relativistic  WSB quark model \cite{wsb} and for $B
\rightarrow S$, $B \rightarrow A$ and $B \rightarrow T$ transitions the ones given in  nonrelativistic  ISGW quark model \cite{isgw} because it is the only quark model that had calculated these transitions. We have considered both $^{1}P_{1}$ and $^{3}P_{1}$ axial-vector mesons. \\

In  second column of Table II, we list $d\Gamma /dt$ for exclusive semileptonic decays $B \rightarrow Ml\nu$ including all $l=0$ and $l=1$ mesons (i.e., pseudoscalar, vector, axial-vector, scalar and tensor mesons) in  final state, in function of form factors (defined in WSB and ISGW quark models) and powers of $\lambda =
\lambda(m_{B}^{2},m_{M}^{2},t)$, where $\lambda=\lambda(x,y,z)=x^{2}+y^{2}+z^{2}-2xy-2xz-2yz$ is the
Euler function and $t = (p_B -p_M)^{2}$ is the momentum transfer. In first row we give this expression for $B \rightarrow Pl\nu$, where $P$ is a pseudoscalar meson \cite{pseudoscalar}; in second row we present $d\Gamma /dt$ for $B \rightarrow Vl\nu$, where $V$ is a vector meson, in function of the form factors $A_{1,2}(t)$ and $V(t)$,  and also in function of  helicity form factors which are defined in  appendix B \cite{vector}; in third row  we show  differential decay width for $B \rightarrow A(^{1}P_{1})l\nu$, where $A(^{1}P_{1})$ is an axial-vector meson \cite{alexander} using the parametrization for  transition $B \rightarrow A$ given in ISGW quark model; in fourth row we display $d\Gamma /dt$ for  $B \rightarrow Sl\nu$. We have obtained this expression from $d\Gamma(B \rightarrow Pl\nu)/dt$ in two steps: first we change the form factors given in WSB quark model by the ones given in ISGW quark model using the formulae showed in appendix A, and second changing the form factors $f_{+,-}$ by  $u_{+,-}$, which are used in the parametrization of $\langle S|J_{\mu}|B \rangle$; in fifth row we present the differential decay rate for $B \rightarrow A(^{3}P_{1})l\nu$, where $A(^{3}P_{1})$ is an axial-vector meson. We have obtained it from the similar expression displayed in third row only changing form factors. Finally, in last row we show $d\Gamma /dt$ for  $B \rightarrow Tl\nu$, where $T$ is a tensor meson \cite{alexander}. \\

\begin{table}[ht]
{\small Table II. Differential decay widths of $B \rightarrow
(P, V, S, A, T)l\nu_{l}$.}
\par
\begin{center}
\renewcommand{\arraystretch}{2}
\begin{tabular}{||c|c||}
  \hline
  Decay & $d\Gamma(B \rightarrow Ml\nu_{l})/dt$ \\
  \hline\hline
  $B \rightarrow Pl\nu_{l}$ & $\zeta \left[A(t)|F_{1}(t)|^{2}\lambda^{3/2} + B(t) |F_{0}(t)|^{2}\lambda^{1/2} \right]$ \\
  \hline
  $B \rightarrow Vl\nu_{l}$ & $\zeta\mathcal{G}(t)$ \\
  \cline{2-2}
   & $\zeta t \lambda^{1/2} \left[ |H_{+}(t)|^{2} + \left|H_{-}(t)\right|^{2}+ |H_{0}(t)|^{2}\right]$ \\
  \hline
  $B \rightarrow A({}^{1}P_{1})l\nu_{l}$ & $\zeta \left\{ \varphi(t)\lambda^{5/2} + \rho(t)\lambda^{3/2}+ \theta(t)\lambda^{1/2} \right\}$ \\
  \hline
  $B \rightarrow Sl\nu_{l}$ & $\zeta \left[A(t)u_{+}^{2}\lambda^{3/2} + B(t) \frac{t^{2}}{(m_{B}^{2}-m_{S}^{2})^{2}}(u_{+}+u_{-})^{2}\lambda^{1/2} \right]$ \\
  \hline
  $B \rightarrow A({}^{3}P_{1})l\nu_{l}$ & $\zeta \left\{ \varphi(t)\lambda^{5/2} + \rho(t)\lambda^{3/2}+ \theta(t)\lambda^{1/2} \right\}$ \\
  \hline
  $B \rightarrow Tl\nu_{l}$ & $\zeta \left\{ \alpha(t)\lambda^{7/2} + \beta(t)\lambda^{5/2}+ \gamma(t)\lambda^{3/2} \right\}$ \\
  \hline
\end{tabular}
\end{center}
\end{table}

In Table II, $F_1(t)$ and $F_0(t)$ are monopolar form factors \cite{wsb},  $u_{\pm}$ are form factors defined in appendix B of Ref. \cite{isgw}. The common factor $\zeta$ and  functions $A(t)$, $B(t)$, $\mathcal{G}(t)$, $\varphi(t)$, $\rho(t)$, $\theta(t)$, $\alpha(t)$, $\beta(t)$ and $\gamma(t)$  are defined by:  \\

\begin{equation}
\zeta = \frac{G_{F}^{2} |V_{qb}|^{2}}{192\pi^{3}m_{B}^{3}},
\end{equation}

\begin{equation}
A(t) = \left( \frac{t-m_{l}^{2}}{t} \right)^{2}
\left(\frac{2t+m_{l}^{2}}{2t} \right),
\end{equation}

\begin{equation}
B(t) = \frac{3}{2} m_{l}^{2} \left( \frac{t-m_{l}^{2}}{t}
\right)^{2} \frac{(m_{B}^{2}- m_{P}^{2})^{2}}{t},
\end{equation}

\begin{align}
\mathcal{G}(t) = \; & \left[
\frac{2t|V(t)|^{2}}{(m_{B}+m_{V})^{2}} +
\frac{(m_{B}+m_{V})^{2} |A_{1}(t)|^{2}}{4m_{V}^{2}}
- \frac{(m_{B}^{2}-m_{V}^{2}-t)
A_{1}(t)A_{2}(t)}{2m_{V}^{2}} \right] \lambda^{3/2}
\nonumber \\
& \;+ \frac{|A_{2}(t)|^{2}}{4m_{V}^{2}
(m_{B}+m_{V})^{2}} \lambda^{5/2} +
3t (m_{B}+m_{V})^{2} |A_{1}(t)|^{2} \lambda^{1/2},
\end{align}

\begin{equation}\label{1}
\varphi(t) = \frac{s_{+}^{2}}{4m_{A}^{2}},
\end{equation}

\begin{equation}\label{2}
\rho(t) = \frac{1}{4m_{A}^{2}}
\left[r^{2}+8m_{A}^{2}tv^{2}+2(m_{B}^{2}-m_{A}^{2}-t)rs_{+}\right],
\end{equation}

\begin{equation}\label{3}
\theta(t) = 3 t\;r^{2},
\end{equation}

\begin{equation}
\alpha(t) = \frac{b_{+}^{2}}{24m_{T}^{4}},
\end{equation}

\begin{equation}
\beta(t) = \frac{1}{24m_{T}^{4}} \left[k^{2}+6m_{T}^{2}th^{2}+2(m_{B}^{2}-m_{T}^{2}-t)kb_{+}\right],
\end{equation}

\begin{equation}
\gamma(t) = \frac{5 tk^{2}}{12m_{T}^{2}},
\end{equation}

where $G_F$ is the Fermi constant, $m_B$ is the mass of $B$ meson, $m_{V(A)}$ is the mass of vector(axial) meson, $m_{P(T)}$ is the mass of  pseudoscalar(tensor) meson,  $m_l$ is the mass of lepton,   $V(t)$ and $A_{1,2}(t)$ are monopolar form factors \cite{wsb}, $\varphi(t)$, $\rho(t)$ and $\theta(t)$ are quadratic
functions of  form factors $s_{+}$, $r$ and $v$ (which are
defined in appendix B of ISGW model \cite{isgw}) for $B \rightarrow A(^{1}P_{1})l\nu$  and of  form factors $c_{+}$, $l$ and $q$ (also given in appendix B of Ref. \cite{isgw}) for  $B \rightarrow A(^{3}P_{1})l\nu$, respectively;  $\alpha(t)$, $\beta(t)$
and $\gamma(t)$ are quadratic functions \cite{alexander} of  form factors $k$,
$b_{+}$ and $h$, which also are given in appendix B of \cite{isgw}.   \\

Comparing  parametrizations of $B \rightarrow P$ and $B \rightarrow S$ transitions given in sections 1 and 5, respectively,  of appendix B of ISGW quark model \cite{isgw} it is easy to relate  expressions of differential decay rates for $B \rightarrow Pl\nu$ and  $B \rightarrow Sl\nu$. In  similar way, if we compare parametrizations of $B \rightarrow V$ and $B \rightarrow A(^{3}P_{1})$ (or $B \rightarrow A(^{1}P_{1})$) transitions given in sections 2 and 4 (or 6), respectively,  of appendix B of ISGW quark model \cite{isgw} it is straightforward to establish a connection between differential decay rates for $B \rightarrow Vl\nu$ and $B \rightarrow A(^{3}P_{1})l\nu$  (or $B \rightarrow A(^{1}P_{1})l\nu$). This is because the role of axial and vector currents is interchanged in both cases. In Ref \cite{calderon} appears a brief discussion about it.     \\

Let us discuss the dependence of $d\Gamma(B \rightarrow Ml\nu)/dt$ with the function  $\lambda$ (note that $|\overrightarrow{p}|=\lambda^{1/2}/2m_B$, where $\overrightarrow{p}$ is the three-momentum of meson $M$ in the $B$ meson rest frame): in general, $d\Gamma/dt \sim \lambda^{l+\frac{1}{2}}$, where $l$ is the orbital angular momentum which is associated to the wave that   particles can be coupled in  final state. Demanding conservation of total angular momentum $J$ and assuming  a meson dominance model it can be found  specific values for $l$ in each exclusive semileptonic $B \rightarrow Ml\nu$ decay. Thus, in $B \rightarrow Pl\nu$ and $B \rightarrow Sl\nu$ particles  are coupled to waves $l=0$ and $l=1$ in final state; in $B \rightarrow Vl\nu$ and $B \rightarrow Al\nu$ particles  can be coupled to waves $l=0$, $l=1$ and $l=2$ in final state; and in $B \rightarrow Tl\nu$ to waves $1$, $2$ and $3$.  \\

We display, in Table III, decay widths of exclusive $W$-external nonleptonic $B \rightarrow M_1,M_2$ decays at tree level (the so called \emph{type-I} decays) assuming factorization hypothesis. We have considered all  the mesons $l=0$ and $l=1$ for $M_1$ and $M_2$. As $\langle T| J_{\mu}|0\rangle = 0$ \cite{gabriel} we do not consider $B \rightarrow M,T$ decays. Again, for transitions $B \rightarrow P$ and $B \rightarrow V$ we use the relativistic WSB quark model \cite{wsb} and the nonrelativistic ISGW quark model \cite{isgw} for $B \rightarrow S$, $B \rightarrow A$ and $B \rightarrow T$ transitions. The WSB quark model only works with $B \rightarrow P$ and  $B \rightarrow V$, i.e., with $l=0$ mesons. For this reason we use both quark models.  Let us mention that in Ref. \cite{deandrea} appears  parametrization of  $B \rightarrow A(^{1}P_{1})$ transition which has the same structure of  parametrization of $B \rightarrow V$ transition interchanging the role of  vector and axial currents: $\langle A| A_{\mu}|B\rangle \leftrightarrow \langle V| V_{\mu}|B\rangle$ and  $\langle A| V_{\mu}|B\rangle \leftrightarrow \langle V| A_{\mu}|B\rangle$. The expressions for $\Gamma(B \rightarrow P_1,P_2$; $P,V$; $V,P$; $V_1,V_2)$ are given in several references (see for example \cite{wsb}); $\Gamma(B \rightarrow T,P$; $T,V)$ were reported in Ref. \cite{tensor} and we have taken $\Gamma(B \rightarrow P,A$; $A,P$; $A,V$; $A,A)$ from Ref. \cite{axial}.\\

\begin{table}[ht]
{\small Table III. Decay widths of $B \rightarrow
P_{1}P_{2},PV,V_{1}V_{2},TP,TV, AP, AV$.}
\par
\begin{center}
\renewcommand{\arraystretch}{1.5}
\begin{tabular}{||c|c||}
  \hline
  Decay & $\Gamma(B \rightarrow M_{1},M_{2})$ \\
  \hline\hline
  $B \rightarrow P_{1},P_{2}$ & $\xi^{(P_{2})}(m_{B}^{2}-m_{1}^{2})^{2} |F_{0}(m^{2}_{2})|^{2} \lambda^{1/2}$ \\
  \hline
  $B \rightarrow P,V$ & $\xi^{(V)}|F_{1}(m^{2}_{V})|^{2} \lambda^{3/2}$ \\
  \hline
  $B \rightarrow V,P$ & $\xi^{(P)}|A_{0}(m^{2}_{P})|^{2} \lambda^{3/2}$ \\
  \hline
  $B \rightarrow V_{1},V_{2}$ & $\xi^{(V_{2})} \mathcal{G}(t=m_{V_2}^{2})$ \\
  \cline{2-2}
   & $\xi^{(V_{2})} m_{2}^{2} \lambda^{1/2} \left[|H_{+}(m^{2}_{V_2})|^{2} + |H_{-}(m^{2}_{V_2})|^{2} + |H_{0}(m^{2}_{V_2})|^{2} \right]$ \\
  \hline
  $B \rightarrow T,P$ & $\xi^{(P)} (1/24m_{T}^{4}) |\mathcal{F}^{B \rightarrow T}(m^{2}_{P})|^{2} \lambda^{5/2}$ \\
  \hline
  $B \rightarrow T,V$ & $\xi^{(V)} \left[ \alpha(m^{2}_{V})\lambda^{7/2} + \beta(m^{2}_{V})\lambda^{5/2}+ \gamma(m^{2}_{V})\lambda^{3/2} \right]$ \\
  \hline
  $B \rightarrow P,A({}^{1}P_{1})$ & $\xi^{(A)}|f_{+}(m_A^{2})|^{2} \lambda^{3/2}$ \\
  \hline
  $B \rightarrow A({}^{1}P_{1}),P$ & $\xi^{(P)}/4m_{A}^{2} \left\{r+s_{+}(m_{B}^{2}-m_{A}^{2}) + s_{-}m_{P}^{2} \right\} \lambda^{3/2}$ \\
  \hline
   $B \rightarrow A({}^{1}P_{1}),V$ & $\xi^{(V)} \left\{ \varphi(m_{V}^{2})\lambda^{5/2} + \rho(m_{V}^{2})\lambda^{3/2}+ \theta(m_{V}^{2})\lambda^{1/2} \right\}$ \\
  \hline
   $B \rightarrow V,A({}^{1}P_{1})$ & $\xi^{(A)} \left\{ \varphi(m_{A}^{2})\lambda^{5/2} + \rho(m_{A}^{2})\lambda^{3/2}+ \theta(m_{A}^{2})\lambda^{1/2} \right\}$ \\
   \hline
\end{tabular}
\end{center}
\end{table}

In Table III, all
 form factors ($F_{1,0}$, $A_{0}$,$A_{1,2}$,$H_{\pm}$, $H_{0}$,$f_+$, $r$, and $s_{\pm}$) are evaluated in $m_2^{2}$ because  the momentum transfer $t = (p_B - p_1)^{2}=p_2^{2}=m_2^{2}$. In similar way    the function $\lambda$ is
 $\lambda(m_{B}^{2},m_{1}^{2},m_{2}^{2})$ for
nonleptonic $B$ decays. $\varphi(m_{A}^{2})$, $\rho(m_{A}^{2})$ and
$\theta(m_{A}^{2})$ has the same form of  Eqs. (\ref{1}),
(\ref{2}) and (\ref{3}) just changing $r \rightarrow f$, $s_{+}
\rightarrow a_{+}$ and $v \rightarrow g$, which are the
appropriate form factors for   $\langle
V|J_{\mu}|B\rangle$ transition in  ISGW-model. We can also add to Table III expressions for
$\Gamma(B \rightarrow P,A({}^{3}P_{1}))$, $\Gamma(B \rightarrow
A({}^{3}P_{1}),P)$, $\Gamma(B \rightarrow A({}^{3}P_{1}),V)$ and
$\Gamma(B \rightarrow V,A({}^{3}P_{1}))$ only changing the form
factors $r \rightarrow l$, $s_{\pm} \rightarrow c_{\pm}$ and $v
\rightarrow q$. The constant
$\xi^{(M_{2})}$ and the function  $\mathcal{F}^{B \rightarrow T}$ are given for the following
expressions:

\begin{equation}
\xi^{(M_{2})} = \;
\frac{G_{F}^{2}|V_{qb}|^{2}|V_{q_{i}q_{j}}|^{2} a_{1}^{2}
f_{M_{2}}^{2}}{32\pi m_{B}^{3}},
\end{equation}

\begin{equation}
\mathcal{F}^{B \rightarrow T}(m_{P}^{2}) = \; k + (m_{B}^{2}-
m_{T}^{2}) b_{+} + m_{P}^{2}b_{-},
\end{equation}

where $k$, $b_{\pm}$ and $h$ are form factors given in  ISGW-model
\cite{isgw},
evaluated at $t = m_{P}^{2}$, $f_{M_2}$ is the decay constant of meson $M_2(q_i\bar{q}_j)$, $a_1$ is the QCD factor,  and $|V_{qb}|$ and $|V_{q_iq_j}|$ are the appropriate CKM factors.\\

Note that in Table III it is straightforward to add decay rates for channels as $B \rightarrow AA$, $B \rightarrow SS$,  $B \rightarrow SP$, $B \rightarrow SV$, $B \rightarrow AS$, $B \rightarrow TA$, and $B \rightarrow TS$  keeping in mind that the role of vector and axial currents of weak interaction is interchanged in $B \rightarrow S$ and $B \rightarrow P$, and in $B \rightarrow V$ and $B \rightarrow A$ transitions.\\

\section{Contributions of vector and axial couplings}

In this section we illustrate how can be coupled particles in  final state in  $B \rightarrow Ml\nu$ and $B \rightarrow M_1M_2$ to specific waves, and determine the quantum numbers of  poles that appear in the monopolar form factors, assuming a meson dominance model.  Moreover, it is possible to check which form factor is related with which wave. We show that these set of circumstances arise from vector and axial couplings of weak interaction. In order to explain this situation we consider the chain $B \rightarrow MM^{*} \rightarrow MW^{*} \rightarrow Ml\nu(q_{1}\bar{q}_{2})$, where $M^{*}$ is the pole and $W^{*}$ is  the off-shell intermediate  boson of weak interaction. We can do it  combining parity and total angular momentum conservations  in  $B \rightarrow MM^{*}$ strong process.   \\

In Table IV, we show if specific waves that  particles can be coupled in final state of semileptonic $B \rightarrow (P,V,S,A,T)l\nu$ decays come from vector or axial contributions. Axial-vector meson $A$ can be $^{1}P_1$ or $^{3}P_1$. In order to explain the respective analysis to each channel we must  keep in mind that the off-shell $W$ boson can  has spin $0$ or $1$. Thus, in vectorial coupling there are two possibilities: $S_W=0$ with $P_W=+1$, and $S_W=1$ with $P_W=-1$ ($S_{W}$ and $P_{W}$ denote spin and parity of the off-shell $W$ boson,  respectively). In  similar way, in  axial coupling there are two options: $S_W=0$ with $P_W=-1$ and $S_W=1$ with $P_W=+1$. Thus, we have four situations: $0^{+}$, $1^{-}$, $0^{-}$ and $1^{+}$. They are displayed in second column of Table IV. Demanding both  total angular momentum conservation ($J_{initial}=J_{final}$) and parity conservation ($P_{initial}=P_{final}$) of the process $B \rightarrow M M^{*}$, where $M^{*}$ is the pole, we found  contributions showed in Table IV. A similar analysis can be performed for  nonleptonic $B \rightarrow M_1M_2$ decays. \\

\begin{table}[ht]
{\small Table IV. Vector and axial contributions to semileptonic $B \rightarrow (P,V,S,A,T)l\nu$ decays.}
\begin{center}
\renewcommand{\arraystretch}{1.3}
\begin{tabular}{||c|c|c|c|c|c|c||}
  \hline
  Contribution & $J^{P}$ of $W^{\ast}$ & $B \rightarrow Pl\nu$ & $B \rightarrow Vl\nu$ & $B \rightarrow Sl\nu$ & $B \rightarrow Al\nu$ & $B \rightarrow Tl\nu$ \\
  \hline\hline
  Vector & $0^{+}$ & $l=0$ &  &  & $l=1$ & \\
  \cline{2-7}
   & $1^{-}$ & $l=1$ &$l=1$  &  & $l=0$, $l=2$&$l=2$\\
  \hline\hline
  Axial & $0^{-}$ &  & $l=1$ &  $l=0$&  & $l=2$\\
  \cline{2-7}
   & $1^{+}$ &  & $l=0$, $l=2$ &$l=1$  & $l=1$  &$l=1$, $l=3$\\
  \hline\hline
\end{tabular}
\end{center}
\end{table}

In Table V, we show the respective form factors  with the corresponding poles in $B \rightarrow Pl\nu$ and $B \rightarrow Vl\nu$ decays. In second column we list the quantum numbers $J^{P}$ of  poles, which are the same $J^{P}$ options for the off-shell $W$ boson (see second column in Table IV). In this case, we must check  form factors that appear in  parametrization of hadronic matrix elements $\langle M| V_{\mu}|B\rangle$ and $\langle M| A_{\mu}|B\rangle$  for $M = P,V$. Following this idea, we can extrapolate quantum numbers of  poles for $B \rightarrow Ml\nu$ where $M$ is a $p$-wave (or orbitally excited) meson: for $B \rightarrow Sl\nu$ the poles are $0^{-}$ and $1^{+}$; for  $B \rightarrow Al\nu$, where axial-vector meson $A$ can be $^{1}P_1$ or $^{3}P_1$, are $0^{+}$, $1^{-}$ and $1^{+}$; and for $B \rightarrow Tl\nu$ the poles are $1^{-}$, $0^{-}$ and  $1^{+}$. This result  is important if we are interested in performing a  quark model with monopolar form factors for  $B \rightarrow S$, $B \rightarrow A$ and $B \rightarrow T$ transitions, i.e., considering $l=1$ mesons in final state.\\

 As an example, we illustrate from Tables IV and V  the situation about $B \rightarrow Pl\nu$: this decay has two contributions $l=0$ and $l=1$ (see the exponents of $\lambda$ in first row of Table II) which arise from  vector coupling of weak current. The respective poles have quantum numbers $0^{+}$ and $1^{-}$, and  form factors are $F_{0}$ and $F_{1}$, respectively (see appendix B). \\

\begin{table}[ht]
{\small Table V. Form factors related to vector and axial contributions of weak interaction to semileptonic  $B \rightarrow (P,V)l\nu$ decays.}
\par
\begin{center}
\renewcommand{\arraystretch}{1.3}
\begin{tabular}{||c|c|c|c||}
  \hline
  Contribution & $J^{P}$ of Pole & $B \rightarrow Pl\nu$ & $B \rightarrow Vl\nu$  \\
  \hline\hline
  Vector & $0^{+}$ & $F_0(t)$&   \\
  \cline{2-4}
   & $1^{-}$ & $F_1(t)$ &$V(t)$\\
  \hline\hline
  Axial & $0^{-}$ &  &$A_0(t)$ \\
  \cline{2-4}
   & $1^{+}$ &  &$A_1(t)$, $A_2(t)$, $A_3(t)$ \\
  \hline\hline
\end{tabular}
\end{center}
\end{table}

\section{Useful ratios}

In this section we establish some ratios using expressions for $d\Gamma(B \rightarrow Ml\nu)/dt$ and $\Gamma(B \rightarrow M_1M_2)$, which are displayed in Tables II and III, respectively. \\

4.1 Let us considerer the \emph{type-I}  $B \rightarrow P^{+,0},V^{-}$ and $B
\rightarrow V^{+,0},P^{-}$ decays, where $P^{+,0}$ and $V^{+,0}$, and,  $V^{-}$ and $P^{-}$ have the same quark content; i.e., the    CKM factors are common in both processes. Moreover, we also assume that   phase spaces are equal. From
Table III, we obtain the following ratio:

\begin{equation}\label{9}
\frac{\Gamma(B \rightarrow P,V)}{\Gamma(B \rightarrow V,P)} =
\left(\frac{f_{V}}{f_{P}} \right)^{2} \left[ \frac{F_{1}^{B
\rightarrow P}(0)}{A_{0}^{B \rightarrow V}(0)}\right]^{2} \left[
\frac{1-m_{P}^{2}/m_{0^{-}}^{2}}{1-m_{V}^{2}/m_{1^{-}}^{2}}\right]^{2},
\end{equation}

where we have used  expressions of  monopolar form factors
$F_{1}$ and $A_{0}$ (see appendix B). As an application of  Eq. (\ref{9}) (following the presentation of Ref. \cite{lu}) we can consider exclusive  $\overline{B^{0}}
\rightarrow \pi^{+},\rho^{-}$ and $\overline{B^{0}} \rightarrow
\rho^{+},\pi^{-}$ decays. Ref. \cite{pdg} reports  $\mathcal{B}(\overline{B^{0}} \rightarrow \rho^{\pm}\pi^{\mp})=(\text{2.28}\pm \text{0.25})$x$10^{-5}$ for the sum of the charge states or particle/antiparticle states indicated. We assume that these branching ratios are approximately equal\footnote{Although we know that the $\rho^{-}\pi^{+}$ decay has a larger rate than the $\rho^{+}\pi^{-}$ mode mainly because of the difference of the decay constants $f_{\rho}$ and $f_{\pi}$ \cite{cheng}.}. Taking  numerical values of masses of  respective mesons reported in \cite{pdg} and pole masses given in Ref. \cite{wsb},  the last factor in  Eq.
(\ref{9}) is approximately 1. Thus, we get:

\begin{equation}\label{10}
1 \approx \frac{\Gamma(\overline{B^{0}} \rightarrow
\pi^{+},\rho^{-})}{\Gamma(\overline{B^{0}} \rightarrow
\rho^{+},\pi^{-})} \approx \left(\frac{f_{\rho^{-}}}{f_{\pi^{-}}}
\right)^{2} \left[ \frac{F_{1}^{B \rightarrow \pi}(0)}{A_{0}^{B
\rightarrow \rho}(0)}\right]^{2}.
\end{equation}

The ratio $\mathcal{R} \equiv F_{1}^{B \rightarrow
\pi}(0)/A_{0}^{B \rightarrow \rho}(0)$ takes different values
according with  theoretical approach used to calculate
numerical values of the form factors evaluated at $t=0$. In
some quark models, for example in \cite{wsb, ebert},  and in \cite{mayoruno}, $\mathcal{R}>1$; from
 lattice QCD calculations \cite{menoruno} and from the light-cone
sum rules (LCSR)  \cite{ball} (Table II of Ref. \cite{cheng}  shows explicitly values of these form factors at zero momentum  transfer obtained in LCSR) is obtained $\mathcal{R}<1$,
and if we take the values of $F_{1}^{B \rightarrow \pi}(0)$ and
$A_{0}^{B \rightarrow \rho}(0)$ reported in Table IV of Ref. \cite{ali}, which are calculated in LCSR \cite{ball} and in lattice-QCD \cite{menoruno}, respectively, we obtain $\mathcal{R} \approx 1$. Thus the
experimental value of $\mathcal{R}$ can discriminate among these
different theoretical approaches\footnote{See for example Refs.  \cite{formfactors, laporta} for a summary about  values of  form
factors at zero momentum transfer $(t=0)$ in different
models.} to evaluate the form factors at $t=0$. If we take the numerical values of decay constants reported in \cite{pdg}: $f_{\pi^{-}}= \text{0.1307}$ GeV  and $f_{\rho^{-}}= \text{0.209}$ GeV, we obtain from Eq. (\ref{10}):

\begin{equation}
\mathcal{R} = \left[ \frac{F_{1}^{B \rightarrow \pi}(0)}{A_{0}^{B
\rightarrow \rho}(0)}\right] \simeq \text{0.632}.
\end{equation}

4.2 Now let us compare  decay widths of $B \rightarrow P,P'$ and $B
\rightarrow P,V$, where $P'$ and $V$ have the same quark content. From  expressions in Table III and using  monopolar form factors (see
appendix B) with the fact that $F_{0}^{B \rightarrow P}(0)=F_{1}^{B \rightarrow P}(0)$, we obtain:

\begin{equation}\label{11}
\frac{\Gamma(B \rightarrow P,P')}{\Gamma(B \rightarrow P,V)} =
\left(\frac{f_{P'}}{f_{V}} \right)^{2} \left[
\frac{1-m_{V}^{2}/m_{1^{-}}^{2}}{1-m_{P'}^{2}/m_{0^{+}}^{2}}\right]^{2}
\frac{\left[\lambda(m_{B}^{2},m_{P}^{2},m_{P'}^{2})\right]^{1/2}}{\left[\lambda(m_{B}^{2},m_{P}^{2},m_{V}^{2})\right]^{3/2}}
(m_{B}^{2}-m_{P}^{2})^{2}.
\end{equation}

Let us considerer the exclusive  $\overline{B^{0}} \rightarrow
\pi^{+},\pi^{-}$ and $\overline{B^{0}} \rightarrow \pi^{+},\rho^{-}$ decays
as an application of  Eq. (\ref{11}). In this case we get:

\begin{equation}\label{12}
\left[\frac{1-m_{\rho}^{2}/m_{1^{-}}^{2}}{1-m_{\pi}^{2}/m_{0^{+}}^{2}}\right]^{2}
\frac{\left[\lambda(m_{B^{0}}^{2},m_{\pi}^{2},m_{\pi}^{2})\right]^{1/2}}{\left[\lambda(m_{B^{0}}^{2},m_{\pi}^{2},m_{\rho}^{2})\right]^{3/2}}
(m_{B^{0}}^{2}-m_{\pi}^{2})^{2} = \text{1.067} \approx 1,
\end{equation}

thus,

\begin{equation}\label{13}
\frac{\Gamma(\overline{B^{0}} \rightarrow
\pi^{+},\pi^{-})}{\Gamma(\overline{B^{0}} \rightarrow
\pi^{+},\rho^{-})} \simeq \left(\frac{f_{\pi^{-}}}{f_{\rho^{-}}}
\right)^{2}.
\end{equation}

This ratio provides only information about  decay constants
$f_{\pi}$ and $f_{\rho}$ and it is independent of  form factors. Using  numerical values of \cite{pdg} $\mathcal{B}(\overline{B^{0}}\rightarrow \pi^{+}\pi^{-})=\text{4.6}$x$10^{-6}$ and $(f_{\pi}/f_{\rho})=(\text{0.1307}/\text{0.209})=\text{0.632}$, we obtain $\mathcal{B}(\overline{B^{0}} \rightarrow
\pi^{+}\rho^{-})=\text{1.15}$x$10^{-5}$. This value is approximately 50.43\% of  $\mathcal{B}(\overline{B^{0}} \rightarrow \rho^{\pm}\pi^{\mp})$ reported in \cite{pdg} for the sum of $\rho^{+}\pi^{-}$ and $\rho^{-}\pi^{+}$ states.\\

We also can apply Eq. (\ref{11}) to decays $B^{0} \rightarrow D^{-},\pi^{+}$ and $B^{0} \rightarrow D^{-},\rho^{+}$ which  branching ratios are $\text{3.4}$x$10^{-3}$ and $\text{7.5}$x$10^{-3}$, respectively \cite{pdg}. Following  a similar procedure in last case, and taking  numerical values of masses of  respective mesons reported in \cite{pdg} and pole masses given in Ref. \cite{wsb} we obtain $(f_{\pi^{+}}/f_{\rho^{+}})=\text{0.651}$.\\

4.3 Another ratio can be obtained  comparing decay widths of $B \rightarrow P,V$ and $B
\rightarrow P,A$, where $V$ and $A$ have the same flavor quantum numbers.  From  expressions in Table III and using  monopolar form factors (see
appendix B)  we obtain:

\begin{equation}\label{15}
\frac{\Gamma(B \rightarrow P,V)}{\Gamma(B \rightarrow P,A)} =
\left(\frac{f_{V}}{f_{A}} \right)^{2} \left[
\frac{1-m_{A}^{2}/m_{1^{-}}^{2}}{1-m_{V}^{2}/m_{1^{-}}^{2}}\right]^{2}
\left[\frac{\lambda(m_{B}^{2},m_{P}^{2},m_{V}^{2})}{\lambda(m_{B}^{2},m_{P}^{2},m_{A}^{2})}\right]^{3/2}.
\end{equation}

We can apply  last equation to  $B^{0} \rightarrow D^{-},\rho^{+}$ and $B^{0} \rightarrow D^{-},a_1^{+}$ decays. From Ref. \cite{pdg} we have $\mathcal{B}(B^{0} \rightarrow D^{-}\rho^{+})=\text{7.5}$x$10^{-3}$ and $\mathcal{B}(B^{0} \rightarrow D^{-}a_1^{+})=\text{6}$x$10^{-3}$. Using these branchings,  numerical values for the respective masses given in \cite{pdg} and the $1^{-}$-pole mass  reported in \cite{wsb}, we obtain from Eq. (\ref{15}) that $(f_{\rho}/f_{a_1})=\text{1.06}$. With $f_{\rho}=\text{0.209}$ GeV \cite{pdg} it is obtained $f_{a_1}=\text{0.197}$ GeV. This value is smaller than the value reported in the literature. For example, Ref. \cite{neubert} gives $f_{a_1}=\text{0.229}$ GeV (extracted from hadronic $\tau$ decay $\tau^{-} \rightarrow M^{-}\nu_{\tau}$) and $f_{a_1}=\text{0.256}$ GeV (comparing branching ratios of $\overline{B^{0}} \rightarrow
D^{*+},a_{1}^{-}$  and $\overline{B^{0}} \rightarrow
D^{*+},\rho^{-}$ decays). On the other hand, Ref. \cite{nardulli} obtained $f_{a_1}=\text{0.215}$ GeV for $\theta = 32^{\circ}$ and $f_{a_1}=\text{0.223}$ GeV for $\theta = 58^{\circ}$, where $\theta$ is the mixing angle between $K_{1A}$ and $K_{1B}$ mesons.\\

We can also relate   $B^{0} \rightarrow \pi^{-},\rho^{+}$ and $B^{0} \rightarrow \pi^{-},a_1^{+}$ decays. Taking $\mathcal{B}(B^{0} \rightarrow \pi^{-}\rho^{+})=\text{1.13}$x$10^{-5}$, $(f_{\rho}/f_{a_1})=\text{1.06}$ (obtained from the before example), the respective numerical values of masses \cite{pdg} and the respective pole mass \cite{wsb}, we obtain from Eq. (\ref{15}):

\begin{equation}\label{16}
\mathcal{B}(B^{0} \rightarrow \pi^{-}a_1^{+}) = \text{9.7x}10^{-6}.
\end{equation}

This prediction satisfies  the upper bound $\mathcal{B}(B^{0} \rightarrow \pi^{\pm}a_1^{\mp}) < \text{4.9}$x$10^{-4}$ reported in \cite{pdg}.\\

4.4 Let us compare $B \rightarrow V,V'$ and $B
\rightarrow V,A$, where $V'$ and $A$ have the same quark content. From Table III we obtained

\begin{equation}\label{17}
\frac{\Gamma(B \rightarrow V,V')}{\Gamma(B \rightarrow V,A)} =
\left(\frac{f_{V'}}{f_{A}} \right)^{2} \frac{\mathcal{G}(m_{V'}^{2})}{\mathcal{G}(m_{A}^{2})}.
\end{equation}

Taking the branching ratios $\mathcal{B}(B^{0} \rightarrow D^{*-}\rho^{+})=\text{6.8}$x$10^{-3}$ and $\mathcal{B}(B^{0} \rightarrow D^{*-}a_1^{+})=\text{1.3}$x$10^{-2}$ \cite{pdg} and evaluating $\mathcal{G}$ using  appropriate monopolar form factors \cite{wsb}, we obtained from Eq. (\ref{17}) that $(f_{\rho}/f_{a_1})=\text{0.81}$, which agrees with the value reported in Ref. \cite{neubert}.  We point out that if we use this value in Eq. (\ref{15}), now we obtain $\mathcal{B}(B^{0} \rightarrow \pi^{-}a_1^{+}) = \text{1.65x}10^{-5}$. This prediction is larger than   the prediction given in  last section (see Eq. (\ref{16})) and it is consistent with experimental upper bound \cite{pdg} for this process. Moreover this prediction is  the same order of   experimental average value $\mathcal{B}(\overline{B^{0}} \rightarrow \pi^{\mp}a_1^{\pm}) = (\text{4.09}\pm \text{0.76})$x$10^{-5}$ \cite{laporta}, which includes BABAR and Belle results,  and theoretical predictions obtained in \cite{calderon,laporta}.\\

4.5 Now, we are going to compare expressions of  decay rate of \emph{type}\emph{-I} nonleptonic $B \rightarrow M_{1},M_{2}$ channel  with  differential decay rate of semileptonic $B \rightarrow M_{1}l\nu_{l}$ process evaluated in $t=m^{2}_{M_{2}}$ at tree level. It is well known that the ratio $\emph{R} = \Gamma(B \rightarrow M_{1},M_{2})/[d\Gamma(B \rightarrow M_{1}l\nu_{l})/dt|_{t=m^{2}_{M_2}}]$ provides a method to test factorization hypothesis and may be used to determine some unknown decay constants \cite{neubert,ratio}. Also it is possible combining exclusive semileptonic and hadronic $B$ decays to measure CKM matrix elements (see for example the paper of J. M. Soares in Ref. \cite{pseudoscalar}). \\

We obtain from Tables II and III, assuming that   $M_2$ is a vector meson  and that   $M_1$ is a $l=0$ or $l=1$ meson:

\begin{equation}\label{18}
\emph{R} = \frac{\Gamma(B \rightarrow M_1,V)}{d\Gamma(B \rightarrow
M_1l\nu_{l})/dt|_{t=m_{V}^{2}}} =
\frac{\xi^{(V)}}{\zeta} =
6\pi^{2}|V_{ij}|^{2}a_{1}^{2}f_{V}^{2},
\end{equation}

where $M_1$ can be a pseudoscalar ($P$) or a vector meson ($V$)  or a scalar ($S$) or an axial-vector ($A$) or a tensor ($T$) meson. $V_{ij}$ is the appropriate CKM matrix element (depending on the flavor quantum numbers of  meson $V$), $f_V$ is the decay constant of  meson $V$, and $a_1$ is the QCD parameter. In general, $\emph{R}=6\pi^{2}|V_{ij}|^{2}a_{1}^{2}f_{V}^{2}X^{(*)}_{M_2}$. So, for  $M_2=V$, one has exactly $X_V =X_V^{*}=X_V^{**}=1$ ($X_V^{**}$ corresponds to the case when $M_1$ is a \emph{p-wave} meson: scalar ($S$), or axial-vector ($A$) or a tensor ($T$) meson).  Thus, the ratio  $\emph{R}$, which is model-independent,  is a clean and direct test of factorization hypothesis. On the other hand, assuming the validity of the factorization with a fixed value for $a_1$, it provides an alternative use: it may be employed for  determination of unknown decay constants.   \\

Now let us to compare the decay width of $B \rightarrow
P_{1},P_{2}$ with the differential semileptonic decay rate of $B
\rightarrow P_{1}l\nu_{l}$ $(m_{l} \approx 0)$. From  Tables II and III, it is obtained:

\begin{equation}\label{19}
\frac{\Gamma(B \rightarrow P_{1},P_{2})}{d\Gamma(B \rightarrow
P_{1}l\nu_{l})/dt|_{t=m_{2}^{2}}} =
\frac{\xi^{(P_{2})}}{\zeta}
\frac{(m_{B}^{2}-m_{1}^{2})^{2}}{\lambda(m_{B}^{2},m_{1}^{2},m_{2}^{2})}
\left[\frac{1-m_{2}^{2}/m_{1^{-}}^{2}}{1-m_{2}^{2}/m_{0^{+}}^{2}}
\right]^{2}.
\end{equation}

If $m_{1}$, $m_{2} \ll m_{B}$, $(m_{B}^{2}-m_{1}^{2})^{2}/\lambda
\approx 1$. This condition also implies that $m_{2} \ll m_{1^{-}},
m_{0^{+}}$ and the last factor in  Eq. (\ref{19}) is of the
order 1. From these conditions, we get:

\begin{equation}\label{20}
\frac{\Gamma(B \rightarrow P_{1},P_{2})}{d\Gamma(B \rightarrow P_{1}l\nu_{l})/dt|_{t=m_{2}^{2}}} \approx \frac{\xi^{(P_{2})}}{\zeta} = 6\pi^{2}|V_{q_{i}q_{j}}|^{2}a_{1}^{2}f_{P_{2}}^{2}.
\end{equation}

As an application of  Eq. (\ref{20}) we can compare   $B \rightarrow \pi,\pi$ and $B \rightarrow \pi l\nu_{l}$ decays. \\

Finally, let us mention that there are several ratios, combining last equations,   that can be of some interest. For example:

 \begin{equation}\label{21}
\frac{\Gamma(B \rightarrow P,V)}{\Gamma(B \rightarrow V_1,V)} = \frac{d\Gamma(B \rightarrow
Pl\nu_{l})/dt|_{t=m_{V}^{2}}}{d\Gamma(B \rightarrow
V_1l\nu_{l})/dt|_{t=m_{V}^{2}}},
\end{equation}

and

\begin{equation}\label{22}
\frac{\Gamma(B \rightarrow P,P_{2})}{d\Gamma(B \rightarrow
Pl\nu)/dt|_{t=m_{2}^{2}}}\frac{d\Gamma(B \rightarrow Pl\nu)/dt|_{t=m_{V}^{2}}}{\Gamma(B \rightarrow P,V)} = \left(\frac{f_{P_2}}{f_V}\right)^{2},
\end{equation}

where $P_2$ and $V$ mesons have the same flavor quantum numbers.

\section{Concluding remarks}

We have performed a brief analysis about  expressions for $\Gamma(B \rightarrow M_1,M_2)$ and $d\Gamma(B \rightarrow M_1l\nu)/dt$ at tree level including all the mesons $l=0$ and $l=1$ in final state. Indeed, we have considered that $M_1$ and $M_2$ can be a pseudoscalar ($P$), a vector ($V$), a scalar ($S$), an axial-vector ($A$) or a tensor ($T$) meson.  We have assumed  factorization hypothesis and used the parametrizations of $\langle M|J_{\mu}|B\rangle$ given in  WSB and ISGW quark models. We explain some aspects related with  dynamics of these processes and give some useful ratios between  decay widths that can determine some form factors, decay constants and branching ratios.

\begin{center}
Acknowledgements
\end{center}

The authors acknowledge financial support from Comit\'e Central de Investigaciones of University of Tolima.

\section*{Appendix A: Relations between form factors of WSB and ISGW quark models.}

 We can  obtain  form factors of WSB model \cite{wsb} in function of  form factors of ISGW model \cite{isgw} comparing the parametrizations given in both models for $B \rightarrow P$ and  $B \rightarrow V$ transitions. Thus, from $\langle P| J_{\mu}|B\rangle_{WSB}$ = $\langle P| J_{\mu}|B\rangle_{ISGW}$ we obtain the following relations:

\begin{equation}
F_{0}(t) = \frac{t}{(m_{B}^{2}-m_{1}^{2})} \;
(f_{+}(t)+f_{-}(t)),
\end{equation}

\begin{equation}
F_{1}(t) = f_{+}(t),
\end{equation}\vspace{0.3cm}

and from $\langle V| J_{\mu}|B\rangle_{WSB}$ = $\langle V| J_{\mu}|B\rangle_{ISGW}$ it is obtained:

\begin{equation}
A_{0}(t) = \frac{1}{2m_{V}} \left[f(t)+ ta_{-}(t)+
(m_{B}^{2}-m_{V}^{2})a_{+}(t) \right],
\end{equation}

\begin{equation}
A_{1}(t) = \frac{f(t)}{(m_{B} + m_{V})},
\end{equation}

\begin{equation}
A_{2}(t) = -(m_{B}+ m_{V}) \;a_{+}(t),
\end{equation}

\begin{equation}
V(t) = -(m_{B}+m_{V})\; g(t).
\end{equation}\vspace{0.5cm}

Using these relations it is straightforward to get $d\Gamma(B \rightarrow P(V)l\nu)/dt$ or $\Gamma(B \rightarrow P(V),M)$ in one model from the respective expressions in the other model.

\section*{Appendix B: Form factors in  WSB quark model.}

In this section we show the expressions for monopolar form factors that appear in the parametrizations of $\langle P|J_{\mu}|B\rangle$  and $\langle V|J_{\mu}|B\rangle$  in WSB model (see Refs.  \cite{wsb} and \cite{korner}):

\begin{equation}
F_{0}(t) = \frac{F_{0}(0)}{1-t/m_{0^{+}}^{2}},
\end{equation}

\begin{equation}
F_{1}(t) = \frac{F_{1}(0)}{1-t/m_{1^{-}}^{2}},
\end{equation}

\begin{equation}
A_{0}(t) = \frac{A_{0}(0)}{1-t/m_{0^{-}}^{2}},
\end{equation}

\begin{equation}
A_{i}(t) = \frac{A_{i}(0)}{1-t/m_{1^{+}}^{2}}, \qquad i=1,2,3.
\end{equation}

\begin{equation}
V(t) = \frac{V(0)}{1-t/m_{1^{-}}^{2}}.
\end{equation}

In section 3 we explain how to obtain  quantum numbers $J^{P}$ for  poles. \\

The helicity form factors are \cite{wsb}:
\begin{equation}
H_{\pm}(t) = (m_{B}+m_{V})A_{1}(t) \mp
\frac{2m_{B}\mathcal{Z}}{(m_{B} + m_{1})} V(t),
\end{equation}

\begin{equation}
H_{0}(t) = \frac{1}{2m_{1}\sqrt{t}} \left[(m_{B}^{2}-
m_{1}^{2}-t)(m_{B} + m_{1})A_{1}(t) -
\frac{4m_{B}^{2}\mathcal{Z}^{2}}{(m_{B}+m_{1})} A_{2}(t)\right],
\end{equation}

where $\mathcal{Z} = 1/2m_{B}\left[(m_{B}^{2}- m_{1}^{2}-t)^{2} -
4m_{1}^{2}t\right]^{1/2}$.

\end{document}